\documentclass[preprint,prb,showpacs,floatfix,superscriptaddress,preprintnumbers,amsmath,amssymb]{revtex4}

\usepackage{graphicx}
\usepackage{dcolumn}
\usepackage{bm}
\usepackage{color}
\usepackage{tabularx}
\usepackage{comment}
\usepackage{latexsym}
\usepackage{amsmath}
\usepackage{amsfonts}
\usepackage{amssymb}
\usepackage{mathtools}
\usepackage{enumitem}
\usepackage{hyperref}
\usepackage{url}
\usepackage{graphicx}
\usepackage{xr}
\usepackage{array}
\usepackage{multirow}

\begin{document}

\newcommand{\grad}{\mbox{$^{\circ}$}}
\newcommand{\cgrad}{\,$^{\circ}$C}
\newcommand{\sixroot}{(6$\sqrt{3}$$\times$6$\sqrt{3}$)R30$^{\circ}$}
\newcommand{\gkdir}{$\overline{\Gamma \mbox{K}}$-direction}
\newcommand{\pb}{$\pi$-band}
\newcommand{\pbs}{$\pi$-bands}
\newcommand{\kv}{\mathbf{k}}
\newcommand{\kpoint}{$\overline{\textrm{K}}$ point}
\newcommand{\kpp}{$\overline{\textrm{K'}}$}
\newcommand{\kp}{$\overline{\textrm{K}}$}
\newcommand{\kps}{$\overline{\textrm{K}_{\textrm{s}}}$}
\newcommand{\kpps}{$\overline{\textrm{K}_{\textrm{s}}\textrm{'}}$}
\newcommand{\km}{$\overline{\textrm{KM}}$}
\newcommand{\gpoint}{$\overline{\textrm{\Gamma}}$ point}
\newcommand{\gk}{$\overline{\Gamma \mbox{K}}$}
\newcommand{\kpar}{$\mathbf{k}_{\parallel}$}
\newcommand{\efermi}{E$_\mathrm{F}$}
\newcommand{\edirac}{E$_\mathrm{D}$}
\newcommand{\angs}{$\mathrm{\AA}$}
\newcommand{\kk}  {$\mathbf{k}_{\overline{\textrm{K}}}$}
\newcommand{\sbs}[2]{\rlap{\textsuperscript{{#1}}}\textsubscript{#2}}
\newcommand{\round}[1]{\left({#1}\right)}
\newcommand{\moire}{moir{\'e}}
\newcommand{\gp}{$\overline{\Gamma}$}
\newcommand{\mpp}{$\overline{\textrm{M}}$}
\newcommand{\kvec}{$\mathbf{k}$}
\newcommand{\qvec}{$\mathbf{q}$}
\newcommand{\mvec}{$\mathbf{m}$}
\newcommand{\gvec}{$\mathbf{g}$}
\newcommand{\svec}{$\mathbf{s}$}
\newcommand{\emini}{E$_\mathrm{m}$}
\newcommand{\red}{\textcolor[rgb]{1,0,0}}
%
%
%
\title{Electronic properties of WS$_2$ on epitaxial graphene on SiC(0001)}

\author{Stiven Forti}%
\affiliation{Center for Nanotechnology Innovation @ NEST, Istituto Italiano di Tecnologia, Piazza San Silvestro 12, 56127 Pisa, Italy}

\author{Antonio Rossi}
\affiliation{Center for Nanotechnology Innovation @ NEST, Istituto Italiano di Tecnologia, Piazza San Silvestro 12, 56127 Pisa, Italy}
\affiliation{NEST, Istituto Nanoscienze-CNR and Scuola Normale Superiore, Piazza S. Silvestro 12, 56127 Pisa (Italy)}

\author{Holger B\"{u}ch}
\affiliation{Center for Nanotechnology Innovation @ NEST, Istituto Italiano di Tecnologia, Piazza San Silvestro 12, 56127 Pisa, Italy}

\author{Tommaso Cavallucci}
\affiliation{NEST, Istituto Nanoscienze-CNR and Scuola Normale Superiore, Piazza S. Silvestro 12, 56127 Pisa (Italy)}

\author{Francesco Bisio}
\affiliation{CNR-SPIN, Corso F. Perrone 24, 16152 Genova, Italy}

\author{Alessandro Sala}
\affiliation{Elettra - Sincrotrone Trieste S.C.p.A., Basovizza, Trieste 34149, Italy}

\author{Tevfik Onur Mente\c{s}}
\affiliation{Elettra - Sincrotrone Trieste S.C.p.A., Basovizza, Trieste 34149, Italy}

\author{Andrea Locatelli}
\affiliation{Elettra - Sincrotrone Trieste S.C.p.A., Basovizza, Trieste 34149, Italy}

\author{Michele Magnozzi}
\affiliation{OPTMATLAB and Dipartimento di Fisica, Universit{\'a} degli Studi di Genova, via Dodecaneso 33 16146, Genova, Italy}

\author{Maurizio Canepa}
\affiliation{OPTMATLAB and Dipartimento di Fisica, Universit{\'a} degli Studi di Genova, via Dodecaneso 33 16146, Genova, Italy}

\author{Kathrin M\"{u}ller}
\affiliation{Max-Planck-Institut f\"{u}r Festk\"{o}rperforschung, Heisenbergstr. 1, D-70569 Stuttgart}

\author{Stefan Link}
\affiliation{Max-Planck-Institut f\"{u}r Festk\"{o}rperforschung, Heisenbergstr. 1, D-70569 Stuttgart}

\author{Ulrich Starke}%
\affiliation{Max-Planck-Institut f\"{u}r Festk\"{o}rperforschung, Heisenbergstr. 1, D-70569 Stuttgart}

\author{Valentina Tozzini}
\affiliation{NEST, Istituto Nanoscienze-CNR and Scuola Normale Superiore, Piazza S. Silvestro 12, 56127 Pisa (Italy)}

\author{Camilla Coletti}
\affiliation{Center for Nanotechnology Innovation @ NEST, Istituto Italiano di Tecnologia, Piazza San Silvestro 12, 56127 Pisa, Italy}
\affiliation{Graphene Labs, Istituto Italiano di Tecnologia, via Morego 30, 16163 Genova, Italy}

\date{\today}

\begin{abstract}
\noindent This work reports an electronic and micro-structural study of an appealing system for optoelectronics: tungsten disulphide WS$_2$ on epitaxial graphene (EG) on SiC(0001). The WS$_2$ is grown via chemical vapor deposition (CVD) onto the EG. Low-energy electron diffraction (LEED) measurements assign the zero-degree orientation as the preferential azimuthal alignment for WS$_2$/EG. The valence-band (VB) structure emerging from this alignment is investigated by means of photoelectron spectroscopy measurements, with both high space and energy resolution. We find that the spin-orbit splitting of monolayer WS$_2$ on graphene is of 462 meV, larger than what is reported to date for other substrates. We determine the value of the work function for the WS$_2$/EG to be 4.5$\pm$0.1 eV. A large shift of the WS$_2$ VB maximum is observed as well , due to the lowering of the WS$_2$ work function caused by the donor-like interfacial states of EG. Density functional theory (DFT) calculations carried out on a coincidence supercell confirm the experimental band structure to an excellent degree.
X-ray photoemission electron microscopy (XPEEM) measurements performed on single WS$_2$ crystals confirm the van der Waals nature of the interface coupling between the two layers. In virtue of its band alignment and large spin-orbit splitting, this system gains strong appeal for optical spin-injection experiments and opto-spintronic applications in general.

\end{abstract}


\maketitle


\section{Introduction}

In recent times, combining two-dimensional (2D) materials with different properties in order to obtain novel van der Waals (vdW) heterostacks with tailored and tunable features\cite{GeimNat2013} has become a possible and tantalizing goal. At present, the most successfully combined 2D materials have been graphene and hexagonal boron nitride (h-BN). The latter provides a great substrate for enhancing graphene's electrical properties and the encapsulation of graphene within h-BN has been proved to be very effective in doing this\cite{MayorovNanoLett2011,BanszerusScience2015}. However, 2D encapsulating layers alternative to h-BN, with better prospects in terms of synthesis and scalability and which might open novel research avenues are being actively seeked for. In this respect, tungsten disulphide (WS$_2$) combined with graphene is a vdW heterostack which hosts a great appeal for applications in optoelectronics. For example, the mobility of graphene encapsulated between WS$_2$ and h-BN is very promising\cite{KretininNanoLett2014}, i.e. about 60000 cm$^2$V$^{-1}$s$^{-1}$. Improving the mobility of graphene by providing an extremely flat substrate and a defect-free interface is only one possible application out of plenty that might emerge.
WS$_2$ has a layer-number dependent band gap and when going from 2 to 1 layer, it exhibits a transition from indirect- to direct-gap semiconductor\cite{KucPRB2011,ZhaoACSNano2013}. The gap in single layer WS$_2$ measures about 2.1 eV \cite{KucPRB2011,HongJPCC2012,JoNanoLett2014} at the two non-equivalent {\kp}-points of its Brillouin Zone (BZ). The neutral exciton in WS$_2$ has a large binding energy\cite{ChernikovPRL2014}, making it a good candidate for the realization of exciton-polariton lasers\cite{FraserNatMat2016}. In virtue of such a long-lived exciton, WS$_2$ shows a remarkably high room-temperature photoluminescence\cite{GutierrezNanoLetters2013}. In the vicinity of the two {\kp} valleys the bands are energy separated because of spin-orbit coupling. The spin-valley coupling is robust enough to observe spontaneous magnetoluminescence at zero magnetic field\cite{ScraceNatNano2015}. Combining semimetallic graphene and semiconducting single-layer WS$_2$ in a vertical heterostack brings together massless Dirac particles with long spin-lifetimes and strongly spin-polarized electrons with great potential for spintronics and optospintronics. Indeed, when placed in close contact, these materials have already shown interesting results in this direction.
The high spin-orbit interaction in WS$_2$ bands has been observed to induce an enhancement of the intrinsic graphene spin-orbit coupling via proximity effect\cite{AvsarNatComm2014}. Moreover, single-layer WS$_2$ was observed to preserve the polarization in photoluminescence experiments\cite{Rossi2DMat2016}. Charge transfer between WS$_2$ and graphene was seen to be fast and efficient under optical pump\cite{HeNatComm2014}. Very recently, a first evidence of tunable spin-injection for stacked flakes of WS$_2$ and graphene has been reported\cite{OmarPRB2017}. The system has therefore a serious appeal for a wide number of applications, ranging from photodetection\cite{TanACSNano2016} to flexible and transparent electronics\cite{GeorgiouNatNano2015}, to optospintronics\cite{GmitraPRB2015}.
However, an in-depth investigation of its electronic properties is still missing.\\
Here we report on the structural and electronic properties of the WS$_2$/graphene system synthesized over large areas via CVD\cite{Rossi2DMat2016}. Investigations are carried out using synchrotron-based X-ray photoemission electron microscopy (XPEEM) for chemical and electronic-structure characterisation, combined with structurally sensitive low-energy electron microscopy\cite{BauerLEEMSpringer} (LEEM). The electronic structure is further probed using angle-resolved photoelectron spectroscopy (ARPES), which provides higher energy resolution. The experimental results are supported by DFT calculations.

\section{Methods}
\subsection{Experimental details}
\label{expdetails}
Nominally on-axis Si-face polished 6H-SiC(0001) purchased from SiCrystal GmbH was used as a substrate for all the experiments. Epitaxial graphene (EG) was grown by thermal decomposition, adapting the recipe of Emtsev and coworkers\cite{Emtsev2009} in an Aixtron Black Magic reaction chamber. WS$_2$ was synthesized by CVD in a hot-wall quartz furnace (cf. also Fig.~\ref{FigS9}), heating up WO$_3$ powder at 900 {\cgrad} for 1 hour and using thermally vaporized sulfur powder as precursor. Argon was used as carrier gas with a flow of 0.5 slm, while the pressure in the reactor was kept at 1 mbar\cite{Rossi2DMat2016}.\\
The microscopy measurements were carried out at the Nanospectroscopy beamline (Elettra Synchrotron, Italy) using the Spectroscopic Photoemission and Low Energy Electron Microscope (SPELEEM) set-up. The SPELEEM combines LEEM with energy-filtered XPEEM. LEEM is a structure-sensitive technique which uses elastically backscattered electrons to image the surface.
In the SPELEEM, a focused, collimated electron beam is generated by a LaB$_6$ gun; the electron energy is precisely set by applying a voltage bias, referred to as start voltage (STV), to the sample\cite{note1}. The lateral resolution of the microscope in LEEM is better than 10 nm\cite{Mentes2014,LocatelliSIA2006}.
Along with imaging, microscopic low-energy electron diffraction ($\mu$LEED) measurements (also known as microprobe-LEED) are performed using illumination apertures to restrict the electron beam to a minimum size of 500 nm.
The SPELEEM is equipped with a bandpass energy filter, allowing to carry out laterally resolved ultra-violet (UV) and soft X-ray photoelectron spectroscopy. In imaging mode, the lateral resolution approaches 30 nm, the energy resolution 300 meV.
XPEEM data at core level energies were evaluated to obtain microscopic photoelectron spectroscopy ($\mu$XPS) spectra. The system is illuminated with photons linearly polarized in the synchrotron's ring plane. The sample is mounted vertical with respect to that plane and the photon beam impinge at 16{\grad} onto the sample. The light is therefore mostly p-polarized. The SPELEEM allows also to carry out microprobe (also known as microspot) ARPES ($\mu$ARPES) measurements\cite{Mentes2012}. With this technique the band structure of the system can be probed on areas as small as $\sim$2 $\mu$m in diameter, allowing the imaging of the angular distribution of photoemitted electrons.

In order to resolve the spin-orbit splitting of the WS$_2$ bands at the {\kp}-point, we carried out ARPES measurements at the Max Planck Institute (MPI) for Solid State Research in Stuttgart. There, ARPES spectra were recorded with a hemispherical SPECS Phoibos 150 electron analyzer in combination with a Scienta VUV5000 lamp. A monochromator selects the He I emission line of the lamp (21.22 eV). 2D dispersion sets E(k) were recorded with the display detector, through a 0.2 mm entrance slit in low angular dispersion mode, corresponding to $\pm$13{\grad} range. With this technique the probed area is of the order of 1 mm$^2$. The mapping of the WS$_2$ BZ was done by measuring single spectra perpendicularly to the high-symmetry direction and varying the photoemission angle. The spectra were acquired at different azimuthal orientations along the {\gp-\kp}, {\gp-\mpp}, {\mpp-\kp} directions. The three different band branches were then put together via software. In this geometrical configuration, the graphene $\pi$-bands intersect the WS$_2$ VB for a small portion. Considering their low cross section at 21 eV and the high emission angle needed close to {\kp}, the $\pi$-bands are not to be seen unless the contrast is strongly enhanced, as in the inset Fig.~\ref{Fig2}(e).

X-ray photoelectron spectroscopy (XPS) spectra were acquired with a Kratos hemispherical analyzer coupled to a monochromatized Al K$_{\alpha}$ X-Ray source. The atomic force microscopy (AFM) images were acquired with a Bruker Dimension Icon microscope used in ScanAsyst tapping mode. Spatially averaged LEED measurements were carried out using an ErLEED system from SPECS GmbH.

All measurements were performed at room temperature.

\subsection{Core-level fitting procedures}
In Sec.\ref{ChemProp} we display the results of local XPEEM measurements. For those measurements, photons at 400 eV were used. For every spectrum a Shirley-type background was considered. We used the reference position of the C 1s peak of SiC in monolayer graphene (MLG) on SiC(0001) (i.e., 283.7 eV\cite{EmtsevPRB2008}), in order to align the binding energy of the spectra extracted for the XPEEM scans. The symmetric peaks were fitted with Voigt functions. To take into account the asymmetry of the peaks coming from conductive layers, such as graphitic carbon, a Doniach-{\v S}unji{\'c} (DS) line shape was used. The C 1s on MLG was fitted taking into account the following components: SiC (Voigt), graphene (DS), S1 (Voigt) and S2 (Voigt), where S1 and S2 are the components associated with the buffer layer\cite{EmtsevPRB2008,RiedlJPD2010,FortiJPD2013}.

\subsection{Computational methods}
The electronic band structure of the system was evaluated within the Density Functional Theory (DFT) framework, using the code Quantum ESPRESSO~\cite{QuantumEspressoPaper}.
The simulation setup is similar to the one we previously used and tested for similar systems\cite{Rippling}. We used a plane wave expansion of the wavefunctions within the pseudopotential~\cite{USPP,RRKJ} approach and a PBEsol~\cite{PBEsol} vdW corrected~\cite{DFT-D2} density functional, both scalar and fully relativistic for spin orbit calculations, using a calculation setup which was previously well tested in similar systems~\cite{Rippling,Antonio}. Both the isolated (i.e. free-standing) $\mathrm{WS_2}$ and $\mathrm{WS_2}$ on top of graphene were studied. To match the graphene and $\mathrm{WS_2}$ lattice parameters a supercell was used, with a periodicity of $(7\times7)$ with respect to $\mathrm{WS_2}$ and $(9\times9)$ with respect to graphene.
At variance with a similar previous calculations\cite{WangNatComm2015} we included vdW correction, which were proven of utmost importance in reproducing inter-layer interactions in graphene-based systems\cite{Rippling}.
The model systems, the supercells and all the simulation setup information are described in detail in the supplementary information (SI).

\section{Results and discussion}
\label{results}

\subsection{Structure and morphology}

\begin{figure}[h!]
\begin{center}
\includegraphics[width=0.5\textwidth]{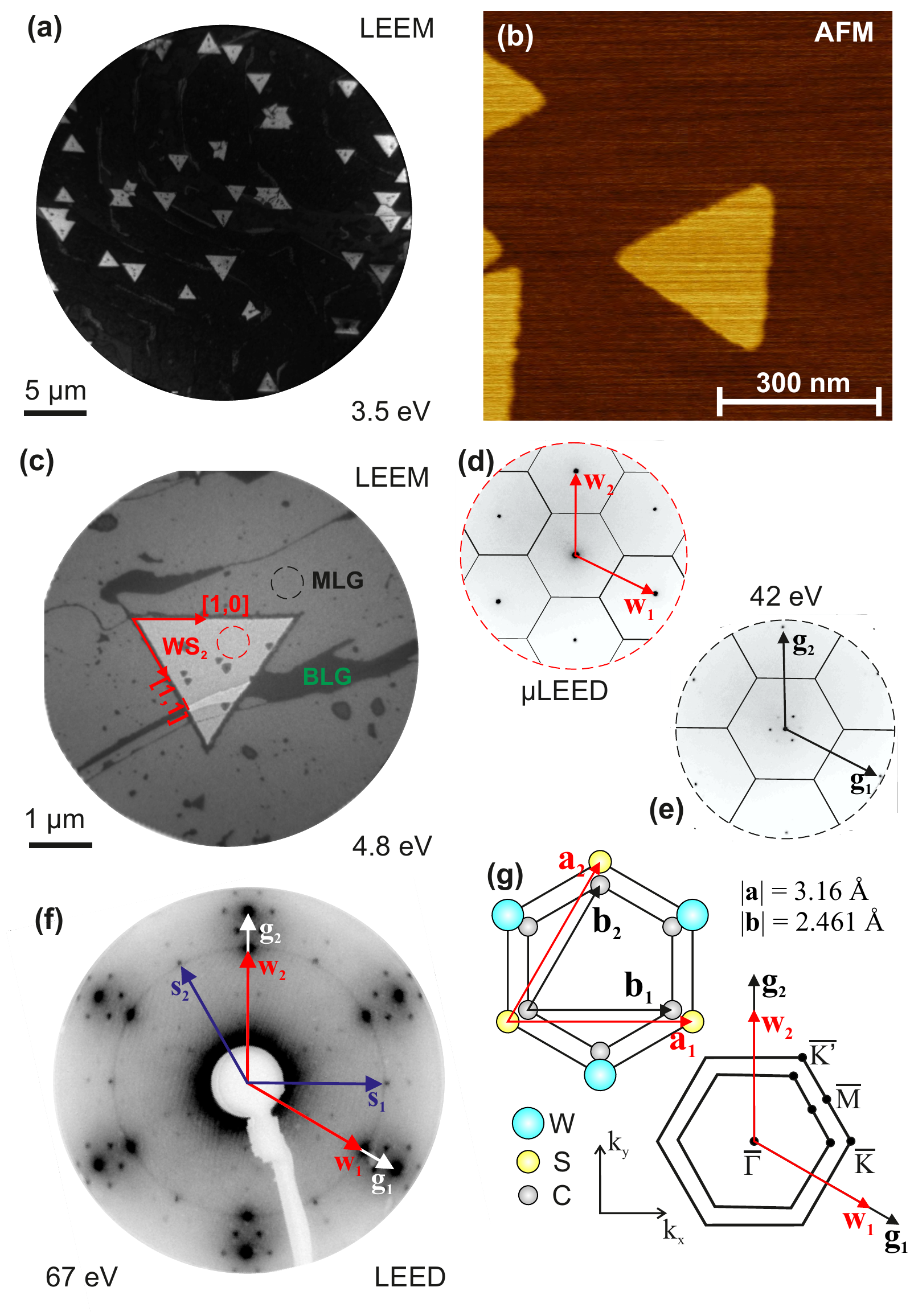}
\end{center}
\caption{(a) LEEM micrograph recorded at STV 3.5 eV on a 30 $\mu$m FOV. (b) AFM image centered on a single WS$_2$ triangle on MLG over a (770$\times$770) nm$^2$ area. (c) LEEM micrograph of a single WS$_2$ crystal imaged at STV 4.8 eV on a 6 $\mu$m FOV. Areas with different contrast are labeled MLG, BLG and WS$_2$, respectively. (d) and (e) $\mu$LEED pattern acquired on the dashed-circle areas for WS$_2$ and MLG, respectively. (f) LEED pattern at 67 eV. WS$_2$, graphene and SiC reciprocal lattice vectors are indicated as $w$, $g$ and $s$, respectively. (g) Real and reciprocal space sketch derived from $\mu$LEED. Crystal directions are drawn in panel (c), according to this sketch.}
\label{Fig1}
\end{figure}

Figure~\ref{Fig1}(a) displays a representative image of the typical $\mu$m-sized triangularly-shaped WS$_2$ single{\--}crystals obtained with our growth approach.
This LEEM micrograph, acquired at STV 3.5 eV over a field of view (FOV) of 30 $\mu$m, indicates that the great majority of the WS$_2$ crystals are aligned along the same crystallographic direction. In Fig.~\ref{Fig1}(b) we show an AFM image with a WS$_2$ triangle to highlight the morphology of the system. The average height of the single-layer WS$_2$ is estimated to be 0.84 nm. The LEEM micrograph in panel (c) is a zoom-in of the image of panel (a) on a single triangle over a 6 $\mu$m FOV. The image was extracted at STV 4.8 eV, hence the different color contrast. $\mu$LEED measurements were performed on the same region, the results of which are shown in panels (d) and (e) for WS$_2$ and graphene, respectively. Because of the very low intensity of the graphene spots when measured on the triangle, we show a measurement acquired outside of the triangle, on a MLG region nearby it. Both measured regions are indicated with dashed circles on the figure. In $\mu$LEED, the (10) spots of WS$_2$ are 3-fold symmetric, resulting from the broken inversion symmetry of the real space lattice, as visible from the sketch in panel (g). In spatially averaged LEED measurements shown in panel (f), the 3-fold symmetry is lost because of the presence of crystals rotated by $n\pi/3$, as visible in panel (a). The preferential alignment of the WS$_2$ along the graphene's crystalline axes is apparent by looking at the (10) diffraction spots of WS$_2$, indicated by $w_{i,j}$ in the figure. The minority orientations are visible as a ring passing through the (10) spots of WS$_2$. The ring is very faint in intensity and its diameter is slightly smaller {\--} about 2.6\% {\--} than the SiC reciprocal lattice vectors ($s_{i,j}$ in the figure). Moreover, the (10) spots of WS$_2$ appear to be slightly elongated along the polar direction, possibly suggesting an equilibrium position fluctuating about the zero degrees orientation. In panel (g) we display the 2D projection of the real and reciprocal space structures of the system as derived from the $\mu$LEED and LEED measurements, assuming the equilibrium value of the graphene's lattice parameter to be 2.461 {\AA} and that of SiC to be 3.08 {\AA}. We find the WS$_2$ lattice parameter to be 3.16$\pm$0.1 {\AA}.
However, no evidence of superperiodicity was found in $\mu$LEED (at higher energies as well), in contrast with WS$_2$/Au(111)\cite{DendzikPRB2015}.
From diffraction measurements we also determine that the edges of the triangular WS$_2$ crystals on epitaxial graphene are aligned along the [1,0] (zigzag) direction, as apparent by comparing panels (c) and (g).

\subsection{Electronic properties}
\label{elec_prop}
The band structure of WS$_2$/MLG was measured by means of $\mu$ARPES on a single WS$_2$ crystal.
\begin{figure*}[ht!]
\begin{center}
\includegraphics[width=0.95\textwidth]{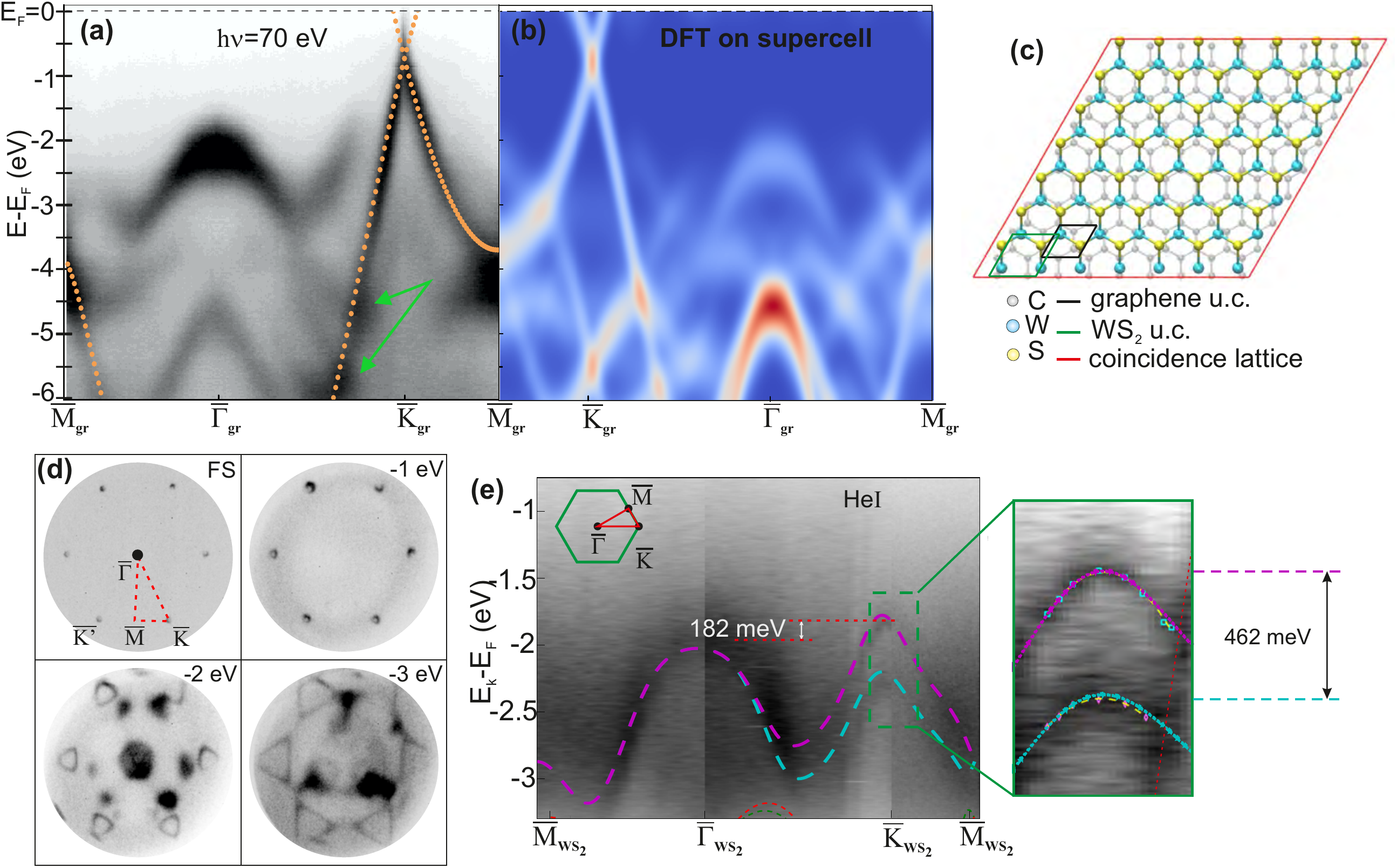}
\end{center}
\caption{Band structure of WS$_2$/MLG. (a) $\mu$ARPES measured on a single WS$_2$ triangle with photons of 70 eV. (b) Theoretical DFT band structure evaluated on the WS$_2$/graphene supercell depicted in panel (c) and unfolded into the graphene's BZ. (c) Coincidence supercell (7$\times$7) over (9$\times$9) of WS$_2$/graphene. (d) Experimental ARPES CESs recorded with  p-polarized photons at 27.5 eV (e) Experimental ARPES band structure of WS$_2$/EG measured with He I light along the path indicated in the inset by the red line. DFT calculated bands including spin-orbit effects are overlapped to the raw data. On the right: zoom-in of the region around {\kp}$_{\text{WS}_2}$ (green-dashed line in panel (e)). Both DFT calculated bands and experimental band fit are overimposed to the raw data. The red-dashed line is on the graphene's {\pb}s.}
\label{Fig2}
\end{figure*}
In particular, the results for the triangle of Fig.~\ref{Fig1}(c) obtained with photons of 70 eV are shown in Fig.~\ref{Fig2}(a). The graphene $\pi$- and $\pi^*$-bands are well visible and also highlighted by orange dots, corresponding to the DFT calculated bands on the graphene single cell. Calculated graphene  $\sigma$-bands are not superimposed as in the experimental data they are not detectable due to their low intensity. The bands visible in  {\gp} belong to WS$_2$ as also indicated by DFT calculations (see Fig.~\ref{FigS4} and Fig.~\ref{FigS5} in the SI). Interestingly, at the points where the bands of graphene and WS$_2$ cross (indicated by green arrows in the panel), no apparent splitting or gap is observed. In order to confirm this finding, DFT calculations were carried out, the result of which is summarized in Fig.~\ref{Fig2}(b), where we display the bands ``unfolded'' onto graphene's BZ, for better readability. The smallest coincidence lattice for the WS$_2$/graphene system was found to be (7$\times$7) WS$_2$ unit cells (u.c.) on (9$\times$9) of graphene, as sketched in Fig.~\ref{Fig2}(c). DFT calculations were carried out on this supercell and as a result, no mini-gap opening was predicted for this system (cf. Fig.~\ref{FigS4} in the SI for further details), contrary to what was recently observed for MoS$_2$ on graphene\cite{PierucciNanoLett2016}. However, the calculations predict that a gap of about 4 meV can be opened at the Dirac point when the distance between WS$_2$ and graphene becomes small enough (see SI), a result which might open interesting scenarios for low-temperature measurements. To better match the experimental data the DFT spectrum was artificially ``doped'' by an energy shift of about 200 meV. The results of the calculations displayed in panel (b) do not include matrix-element effects depending on the incident photons and thus the intensity distribution cannot be directly compared with the experimental ARPES data. To better visualize the relation between the graphene {\pb}s and the WS$_2$ bands, we show constant energy surfaces (CESs) in Fig.~\ref{Fig2}(d), extracted starting from the Fermi surface (FS) at binding energies indicated in the figure. The CESs in this case are small volumes in $k$ space integrated over about 250 meV, corresponding to the resolution of the instrument. The data were acquired with a photon energy of 27.5 eV in order to maximize the intensity of the WS$_2$  bands with respect to graphene (see Fig~\ref{FigS5}).

We display the results of the ARPES measurements recorded at the MPI with He I radiation in Fig.~\ref{Fig2}(e), together with the DFT-calculated bands including spin-orbit coupling. The image was obtained by scanning the BZ of the system along the red line traced within the green hexagon in the inset (cf. Fig.~\ref{FigS5}(d) of the SI for the raw data). Note that in this image the high symmetry points are for the WS$_2$ BZ, whereas for panels (a) and (b) we referred to graphene's BZ. Single spectra were measured perpendicular to the red line.

We have fitted the experimental data in proximity ($\pm\sim0.1$ \AA$^{-1}$) of {\kp} with a parabolic function in order to extract the effective mass values of the holes.
The result along the {\gp}-{\kp}-{\mpp} direction is displayed on the right side of Fig.~\ref{Fig2}(e), representing the zoom-in of the region framed with a green-dashed line in the panel. We find $m_{h1}\simeq0.39m_e$ for the low energy band and $m_{h2}\simeq0.53m_e$ for the high energy band, confirming the asymmetry reported in other publications\cite{DendzikPRB2015}.

The spin-orbit splitting of the WS$_2$ bands in {\kp} was retrieved from integrated energy distribution curves (EDCs) to be 462 $\pm$ 5 meV (see also Fig.~\ref{FigS7} in the SI). Notably, this value is about 10\% larger than what was measured for monolayer WS$_2$ on Au(111) and Ag(111) \cite{DendzikPRB2015,UlstrupPRB2017} and about 7\% larger than the highest value reported so far\cite{KatochArxiv2017}. The value measured on our system is comparable only with measurements carried out on bulk WS$_2$\cite{YuanNanoLett2016}.

\begin{figure*}[h!]
\begin{center}
\includegraphics[width=0.95\textwidth]{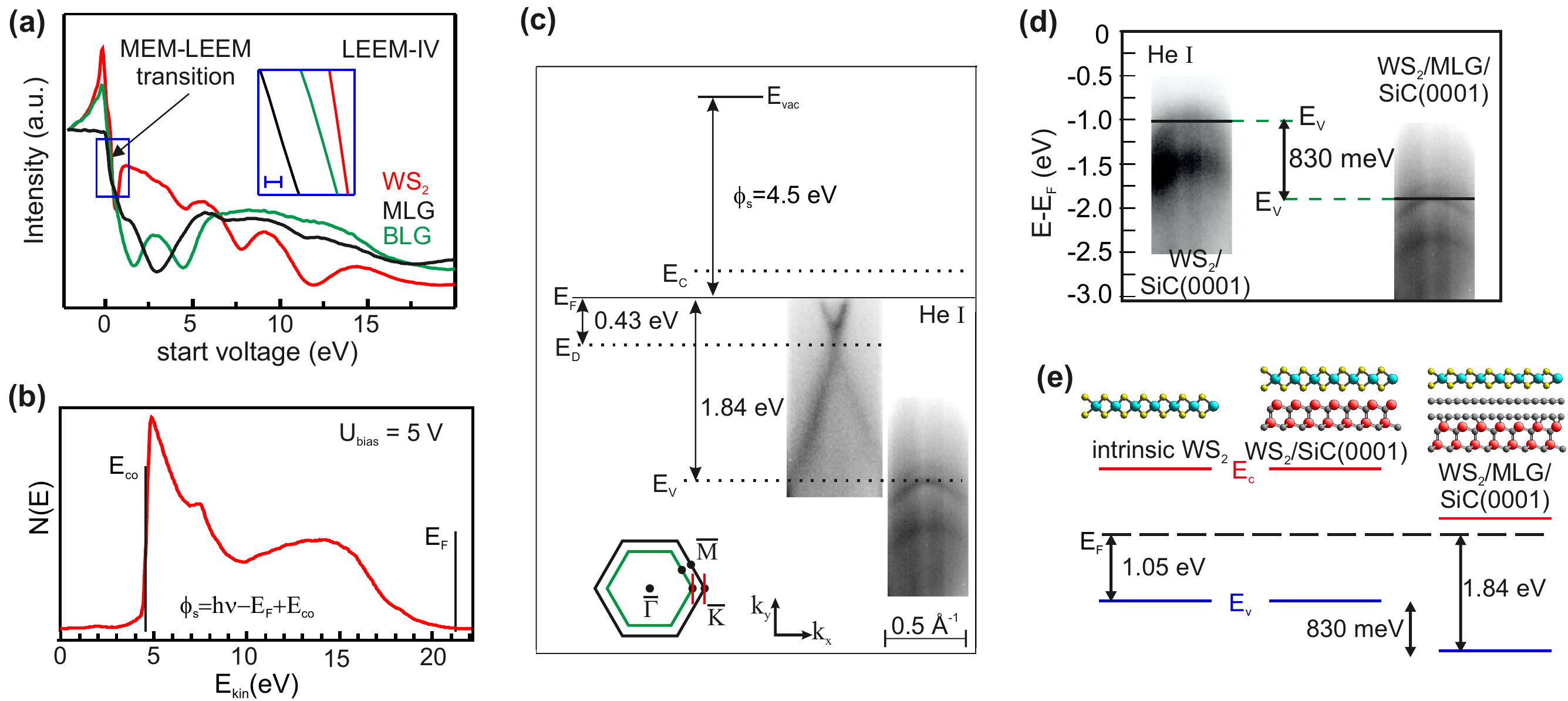}
\end{center}
\caption{(a) LEEM-IV curves measured on and outside the triangle on the MLG, BLG and WS$_2$ regions defined in Fig.~\ref{Fig1}(c). In the inset, the bar is 50 meV. (b) Experimental UPS data acquired with a 5 V bias voltage applied to the sample. (c) Scheme of the band alignment derived from the ARPES spectra measured with He I light along the red lines shown in the inset. (d) On the left, bands of WS$_2$/SiC(0001) and on the right WS$_2$/MLG/SiC(0001). Both measurements are aligned to the Fermi energy and the energy scale is the same. (e) Simple scheme of the different energy alignments for intrinsic (ideal) WS$_2$, WS$_2$/SiC(0001) and WS$_2$/MLG. On the top, a schematic representation of each considered system.}
\label{Fig3}
\end{figure*}

In Fig.~\ref{Fig3}(a) we show LEEM-IV spectra recorded on WS$_2$, MLG and bilayer graphene (BLG) areas, as labeled in Fig.~\ref{Fig1}(c).
LEEM-IV curves give information about the electronic properties of the system for energies above {\efermi} and their dips, at least in the case of graphene, indicate the number of layers\cite{HibinoPRB2008,SrivastavaPRB2013,FeenstraUM2013,FeenstraPRB2013}. In the WS$_2$ spectrum we observe three dips modulated by a linear decay of the intensity, possibly reflecting the three-layer structure of the  single-layer WS$_2$.\\
LEEM-IV curves can also provide a direct and local measurement of the surface potential difference between different regions looking at the transition between mirror mode (MEM) and LEEM\cite{Mentes2014}. In the inset of the figure we show a zoom-in of the MEM-LEEM transition region with an energy scale-bar of 50 meV. We observe that the WS$_2$/MLG exhibits a value of work function slightly larger (about 150 meV) than of pristine (or as grown) MLG. The value of the work function instead, was obtained over the entire sample from HeI UV photoelectron spectroscopy (UPS) measurements of the VB. The VB spectrum is shown in panel (b) and was acquired with the sample biased at 5 V in order to access the secondary electrons cut-off energy. The work function of the analyzer is constant and the acquisition software compensate for it in a way that the kinetic energy of the electrons at the Fermi level, essentially coincide with the photon energy. The work function of the sample $\phi_S$ is then $h\nu-${\efermi}$+E_{co}$, where $E_{co}$  is the cut-off energy beyond which no electron is emitted from the sample. The value obtained in this way is 4.35 $\pm$ 0.05 eV.
By combining these information we could determine the band alignment of the system, which is displayed in Fig.~\ref{Fig3}(c).

Although the role of the substrate requires further investigation, in Fig.~\ref{Fig3}(d) we provide a first proof of its relevance. On the left side we show the ARPES spectrum of WS$_2$ grown directly on 6H-SiC(0001) following the procedure described in section~\ref{expdetails}. The bands are recorded in {\kp} in the same geometry as shown in the inset of panel (a). On the right side, we show again the ARPES of WS$_2$/MLG as in panel (c). The energy difference between the two valence band maxima (E$_v$) is about 830 meV. This indicates that the WS$_2$, when grown on SiC(0001), exhibits a band alignment very close to the one expected for isolated WS$_2$, i.e. Fermi level in the middle of the bandgap.
For the size of the bandgap we refere here to recent time-resolved ARPES measurements\cite{UlstrupPRB2017}, which set the bandgap value for single layer WS$_2$ at 2.1 eV.
To make this visually more clear, in panel (e) we show a scheme of the band alignment for the three situations: isolated WS$_2$, WS$_2$/SiC(0001) and WS$_2$/MLG. On top of every sketch, a simple ball-stick model of each system is shown.
For the WS$_2$/MLG, the conduction band minimum (E$_c$) lays then $\sim$260 meV above the Fermi level. In the case of WS$_2$/SiC(0001) instead, the valence band maximum (E$_v$) is found at 1.00$\pm$0.05 eV below {\efermi}, which means that the donor states of the graphene/buffer layer system\cite{RisteinPRL2012,KopylovAPL2010} ``pin'' the Fermi level of WS$_2$ on MLG, thereby lowering its work function. The spin-orbit splitting of the bands in {\kp} remains instead unaltered. Indeed, for the WS$_2$/SiC(0001), the $\Delta$E between the fitted maxima of the peaks of the integrated intensity yields $458\pm5$ meV.

The population of the conduction band via transfer of negative charge to TMDs leads to unconventional phenomena as negative electronic compressibility (NEC), as observed in WSe$_2$\cite{RileyNatNano2015} and more recently also in WS$_2$\cite{KatochArxiv2017}. The NEC reduces the size of the gap and since bilayer WS$_2$ has a smaller band-gap, it could be readily metallic on EG, opening up the possibility for the observation of predicted exotic phenomena such as the transition to a superconductive phase\cite{JoNanoLett2015,ZhangNanoLett2016}. In addition, the energy difference between the maxima of the VB in {\gp} and {\kp} is found to be $\Delta_{\Gamma K}=$182 meV (cf. Fig.~\ref{Fig2}(e)), about a third of what was observed for the same material on Au(111)\cite{DendzikPRB2015}, possibly implying the occurrence of many-body renormalization effects of the bands or due to the graphene-WS$_2$ interaction. As a comparison, we note that for the WS$_2$/SiC(0001) system the same quantity was found to be $\Delta_{\Gamma K}=(250\pm20)$ meV (not shown).

Despite the recent popularity of vdW vertical heterostructures and the variety of investigated TMDs, the system presented and studied in this work represents an \textit{unicum} as referred to the potential applications in opto-spintronics. Graphene has a large spin relaxation time\cite{TombrosNat2007}, but the electrical injection of spin in graphene suffers of problems arising from the quality of the contacts, defects at the interfaces, minority spin injection or the definition of a tunnel barrier to minimize it\cite{WuPRA2014}. A cleaner way to inject spin polarized carriers into graphene would be optically, i.e. by exploiting the optical selection rules introduced by the use of photons with a specific helicity. We propose that the WS$_2$/MLG described in this work has the ideal band alignment for such applications using photons in the visible range.

\subsection{Chemical properties}
\label{ChemProp}
\begin{figure}
\begin{center}
\includegraphics[width=0.5\textwidth]{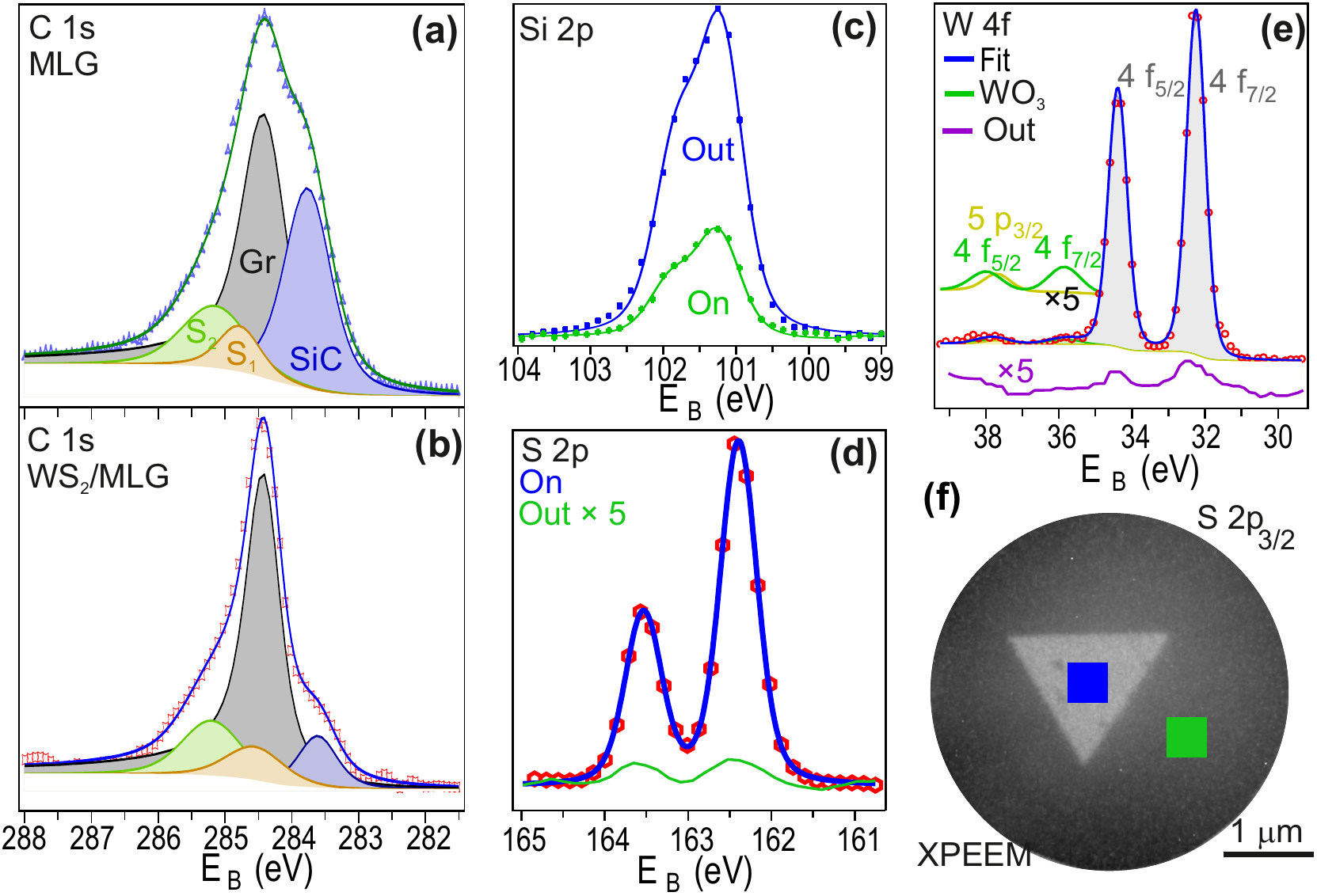}
\end{center}
\caption{Panels (a-d): Core level photoemission spectroscopy spectra extracted from XPEEM scan at 400 eV.  The label ``out'' indicates that the spectrum was recorded outside the triangle. (f) XPEEM snapshot of S 2p extracted on the high-spin component highlighting the areas from which the spectra are extracted. The same areas are valid for the W 4f spectra.}
\label{Fig5}
\end{figure}
The chemical properties of the system are investigated locally, i.e. measuring each core level spectrum on and outside a single WS$_2$ crystal, by means of XPEEM. The outcome of those measurements are summarized in Fig.~\ref{Fig5}(a-e).
Laterally averaged chemical properties were instead collected via XPS, as described in the methods section and the data are summarized in Fig.~\ref{FigS6} of the SI.

In Fig.~\ref{Fig5}(a) and (b) we compare the C 1s spectra recorded on MLG and WS$_2$/MLG, respectively. Intensities are area normalized so that line-shape and peak positions can be better compared. We point out that, within the experimental error, the positions of the C 1s components do not shift. In particular, the sp$^2$ graphitic peak remains at 284.4$\pm$0.1 eV, confirming the absence of doping variation in graphene (cf. section~\ref{elec_prop}) and at the same time excluding strong chemical interaction between the two 2D layers. Because their shape is not explicitly evident at this particular photon energy, the S1 and S2 components {\--} characterizing the buffer layer {\--} were assigned from literature data\cite{EmtsevPRB2008}. We find S1 at 284.7$\pm$0.2 eV and S2 at 285.2$\pm$0.1 eV. Again, their positions are stable on and outside the WS$_2$ island. The SiC component is found at 283.7$\pm$0.1 eV in both cases, meaning that the WS$_2$ layer does not induce any band bending of the SiC core-level bands. Such a fact is further confirmed by the Si 2p peak, shown in Fig.\ref{Fig5}(c). Also in that case, the 2p doublet remains at 101.2$\pm$0.1 eV.
Panels (d) and (e) display the S 2p and W 4f spectra, respectively. The S 2p is well fitted with a single Voigt doublet with Lorentian width 0.09 eV and Gaussian width 0.2 eV, with the 2p$_{3/2}$ component centered at 262.2$\pm$0.1 eV. This is symptomatic of the fact that the sulphur atoms of both top and bottom layers are in the same chemical environment and the interaction with the graphene p$_z$ orbitals does not induce a measurable chemical shift.
The W 4f contains a visible second component that we ascribe to a high-oxidation state, namely WO$_3$. The intensity of the oxide doublet is about 4.5\% of that of the 4f disulphide doublet. The energy position of the WS$_2$ W 4f$_{7/2}$ component is measured as 32.2$\pm$0.1 eV and the spin-orbit splitting 2.15$\pm$0.05 eV.
As for the other peaks, we measured W 4f and S 2p also outside the WS$_2$ triangle and we report those spectra with intensity multiplied by a factor 5 in the figures. We find some sulphur and tungsten with energies compatible with those of WS$_2$. The WO$_3$ was instead detected only on the island, leading us to the conclusion that some unreacted material is embedded into the WS$_2$ or underneath it. In panel (f) we display the XPEEM snapshot acquired at the S 2p$_{3/2}$ energy, showing the regions where the spectra in and outside the triangle were acquired from.

\section{Conclusions}
In this article we investigate the properties of CVD-grown WS$_2$ crystals on epitaxial graphene on SiC(0001) by means of microscopy techniques such as LEEM/PEEM and AFM as well as laterally averaging methods as ARPES and XPS.
The set of measurements we carry out converge on defining the WS$_2$/MLG a low-interacting system. Indeed, $\mu$LEED does not show {\moire}-like diffraction spots and neither $\mu$ARPES nor ARPES show replica bands. DFT calculations support the experimental findings by evidencing the absence of gaps (either due to band anticrossing or to superperiodicity effects), which have instead been recently reported for a similar system, e.g. MoS$_2$/graphene.
The analysis of core level data excludes substantial chemical shifts and line-shape modifications such as peak broadening or splitting, further confirming the weak interaction between WS$_2$ and graphene. The band alignment between WS$_2$ and graphene is determined. We find that the position of the MLG Dirac point does not change, behavior observed also for MoS$_2$ on MLG\cite{UlstrupACSNano2016}, and the E$_v$ of WS$_2$ is located 1.84 eV below {\efermi}. This strong downshift of about 830 meV of the VB maximum depends on the substrate and it alters the value of work function for the WS$_2$. The band structure of the system is measured through the entire BZ. We extract the effective masses in {\kp}, finding $m_{h1}\simeq0.39m_e$ for the low energy band and $m_{h2}\simeq0.53m_e$ for the high energy band. The spin-orbit splitting of the VB at {\kp} is found to be 462 meV, the highest values reported for this material in its monolayer form. Together with the observed 0{\grad} azimuthal alignment of the two crystals, the band structure of the system results to be promising for applications in the realm of opto-, spin- and valleytronics.

\section*{Acknowledgments}
The research leading to these results has received funding from the European Union
Seventh Framework Program under grant agreement no. 604391 core 1 Graphene Flagship.
We gratefully acknowledge CINECA for providing  HPC resources under the ISCRA-C grants ``Quasi free standing graphene monolayer on SiC with H-coverage vacancies: a density functional theory study'' (2016-2017), and ``Electro-mechanical manipulation of graphene'' (2015-2016), and for technical support. This work was partly supported by the German Research Foundation (DFG) in the framework of the Priority Program 1459 Graphene.
The authors also thank the PRIN project 20105ZZTSE{\_}003 for financial support. We thank Alberto Morgante for useful discussions.




\renewcommand\refname{Notes and references}


\begin{thebibliography}{10}
\bibitem{note1}
The actual electron energy can be defined by performing a LEEM-IV scan. In
  particular, the electron energy is defined as the start volage minus the
  start voltage of the MEM-LEEM transition. Cf. Fig.~\ref{Fig3}(a).
\input{note1}
\end{thebibliography}

\end{document}